\documentclass{article}

\PassOptionsToPackage{numbers, square}{natbib}
 \usepackage[preprint]{neurips_2026}

\usepackage[utf8]{inputenc} 
\usepackage[T1]{fontenc}    
\usepackage{lmodern}        
\usepackage[colorlinks=true, linkcolor=blue, citecolor=blue]{hyperref} 
\usepackage{url}            
\usepackage{booktabs}       
\usepackage{amsfonts}       
\usepackage{nicefrac}       
\usepackage{microtype}      
\usepackage{xcolor}         
\usepackage{listings}
\usepackage{amsmath}
\usepackage[font=small]{caption}
\usepackage{graphicx}
\graphicspath{ {./images/} }

\lstdefinestyle{mystyle}{
    basicstyle=\ttfamily\small,
    keywordstyle=\bfseries\color{blue!70!black},
    commentstyle=\itshape\color{green!50!black},
    stringstyle=\color{red!70!black},
    numberstyle=\tiny\color{gray},
    identifierstyle=\color{black},
    emphstyle=\color{purple!70!black},
    numbers=none,
    breaklines=true,
    showstringspaces=false,
    tabsize=4,
    backgroundcolor=\color{gray!8},
    frame=single,
    rulecolor=\color{gray!30},
    framexleftmargin=5pt,
    framexrightmargin=5pt,
    framextopmargin=3pt,
    framexbottommargin=3pt,
    framerule=0.4pt,
    xleftmargin=8pt,
    xrightmargin=8pt,
    morekeywords={void,int,uint,undefined4,longlong,return,if,else}
}
\title{Identifier-Free Code Embedding Models for Scalable Search}

\author{%
  Eric Wolos\thanks{corresponding author} \\
  The MITRE Corporation\\
  San Diego, CA 92106 \\
  \texttt{ewolos@mitre.org} \\
  \And
  Michael Doyle \\
  The MITRE Corporation\\
  San Diego, CA 92106 \\
  \texttt{mdoyle@mitre.org} \\
}

\begin{document}

\maketitle
\setcounter{footnote}{0}
\renewcommand{\thefootnote}{\arabic{footnote}}

\begin{abstract}
  Function association is a useful process for binary reverse engineers. Search tools exist to perform association at scale, but they do not utilize the full range of capabilities that AI-enabled search provides. Prior work has explored the development of embedding models for association between certain reverse engineering code representations, but that work does not cover bidirectional association between source code and decompiled, stripped code with standard preprocessing requirements. To bridge this gap, we formalize this function association problem and evaluate the extent to which embedding models can bidirectionally associate between these two representations. To improve model performance at this task, we fine-tune a Qwen3-Embedding model with contrastive learning. We find that our new model outperforms other models on all function association baselines by a substantial margin and generalizes to a constant-algorithm association task it is not explicitly trained on.

\renewcommand\thefootnote{}\footnotetext{Copyright © 2026 The MITRE Corporation. ALL RIGHTS RESERVED. Approved for Public Release; Distribution Unlimited. Public Release Case Number 26-0893}

\end{abstract}

\section{Introduction}

\subsection{Background}

Binary reverse engineering involves reasoning about how a particular binary works. Since many binaries are compiled with debug and symbol information stripped, binary reverse engineers frequently have little naming information for understanding the code they analyze. Using binary reverse engineering tools like Ghidra\footnote{\url{https://ghidra-sre.org/}} and IDA Pro\footnote{\url{https://hex-rays.com/ida-pro/}}, engineers can see pseudo-C representations of the original source code. While these estimations usually produce functionally equivalent functions, identifier information about variable, parameter, and function names is lost and control flow may be rearranged.

One strategy reverse engineers may use during their analysis is to associate functions of interest to other functions that they already understand. Function association can be accomplished with existing search tools like BSim \citep{BSim} or FLIRT \citep{flirt}. Such rules-based methods are effective and useful, but may be supplemented by AI-enabled function association.

\subsection{Related Works}

Improving general code embedding has been an important problem in embedding-based search. The use of deep learning to associate code representations across binaries has also seen significant recent exploration. \citet{word2vecmal} has demonstrated the efficacy of a Word2Vec variant on detecting malware. For general function association, \citet{10.1145/3728927} and \citet{chen2026crossmodalretrievalmodelsstripped} study retrieving binary code via natural language with various architectures. To observe potential generalization of such models, we compare BinSeek-Embedding from \citet{chen2026crossmodalretrievalmodelsstripped} alongside ours, although we acknowledge its natural language-based purpose and training objective differ from ours. Others, like \citet{Wang_Zhou_Han_Zhang_2026}, perform cross-architecture associations between binaries, allowing for search between binaries compiled for different architectures. General embeddings for assembly language and lifted (disassembly) languages exist as well, and many prior works leverage these representations for source code association, such as \citet{zhang2025pretrainingrepresentationsbinarycode}, \citet{Li_2021}, and \citet{Jia2024SemanticAwareInstructionEmbedding}. To map source code and corresponding Control Flow Graphs, \citet{NEURIPS2020_285f89b8} use convolutional and graph neural networks, while requiring feature engineering for string / integer literal extraction. Finally, \citet{YU2025129238} approach code associations across source and decompiled, stripped code with LSTM networks but require nonstandard preprocessing such as building abstract path context (APC) data and abstract syntax trees. This processing requires tools like Ctags\footnote{\url{https://ctags.io/}}, Cscope\footnote{\url{https://cscope.sourceforge.net/}}, IDA Pro, and ASTMiner\footnote{\url{https://github.com/JetBrains-Research/astminer}} and its downstream model is not openly accessible.

\subsection{Motivation}

We seek to examine search in both directions between source code and decompiled, stripped code representations. To make this search transferable to applications, we use transformer-based embedding models with text inputs to avoid sophisticated input preprocessing. Next, we train a model with contrastive learning to measure improvement on this task. In addition, we seek to measure model generalization on related tasks with objectives that models are not explicitly trained on.

This work focuses on source code and the symbol-stripped, pseudo-C, decompiled code that tools like Ghidra and IDA Pro generate (see Function Example 1). We focus on text embedding transformers because they allow us to relax constraints on model inputs to any text that can be tokenized. Not only does this keep preprocessing simple, but it also allows analysts to easily write their own query functions and use code that is not currently compiled. This work does not seek to solve the function association problem as a whole, but rather examine a methodology for mapping particular representations and understand how our training paradigm affects underlying model performance at highly specific, data-constrained tasks.

\section{Methods}

\subsection{Datasets}

\subsubsection{Assemblage}

Assemblage \citep{liu2024assemblageautomaticbinarydataset} is a large dataset of binaries written in C/C++ and compiled for x86 via an automated scraping and compilation pipeline. Originally taken from GitHub, Assemblage's WinPE binaries are the only portion of the dataset with direct source code attribution, so we only use these in our work. All references to Assemblage in this work specifically refer to the "permissively-licensed" portion of this split. We extend the Assemblage dataset by decompiling its binaries with headless Ghidra. 

Our fraction of the Assemblage dataset contains nearly 500,000 function pairs (over 1M total functions across both representations), split approximately 95\%/5\% for training and testing. Validation data is held out from the training split, as described in Section 3.1. We leave all original comments in the source code intact.

To give a concrete example of these two representations, we present the same \textit{Function Example 1} \citep{imgui} contained in Assemblage in both source and decompiled forms. Note that increases in the number of function parameters are expected because of the \textit{this} pointer being passed implicitly by C++ calling convention in non-static member functions. Identifiers like function and variable names in the decompiled example are determined during decompilation by Ghidra itself, since all names were stripped in the binary as part of the compiling process. 

\vspace{0.5em}
\noindent\textit{Function Example 1:}
\vspace{0.3em}

\small
{
\noindent\textbf{Source Code} 
\begin{lstlisting}[language=C] 
void ImGuiStorage::SetInt(ImGuiID key, int val)
{
    ImGuiStoragePair* it = LowerBound(Data, key);
    if (it == Data.end() || it->key != key)
    {
        Data.insert(it, ImGuiStoragePair(key, val));
        return;
    }
    it->val_i = val;
}

\end{lstlisting}

\textbf{Decompiled}
\begin{lstlisting}[language=C]
void FUN_140006198(int *param_1,uint param_2,uint param_3)

{
  uint *puVar1;
  uint *puVar2;
  undefined4 *puVar3;
  undefined4 local_18 [4];
  
  puVar1 = (uint *)FUN_1400060a8(param_1,param_2);
  puVar2 = (uint *)FUN_14001c324(param_1);
  if ((puVar1 == puVar2) || (*puVar1 != param_2)) {
    puVar3 = FUN_140002d74(local_18,param_2,param_3);
    FUN_14001c2a0(param_1,(longlong)puVar1,puVar3);
  }
  else {
    puVar1[2] = param_3;
  }
  return;
}
\end{lstlisting}
}

\subsubsection{Signsrch}

Signsrch, developed by \citet{signsrch} (GPL license), is a tool for identifying various cryptographic, compression, and multimedia algorithms by automatically searching file signatures for special constant values. Signsrch is designed to identify usage of proprietary filetypes or protocols within binaries. An analyst might similarly use the tool's underlying constant data to manually identify algorithms during reverse engineering.

Example algorithms include forms of CRC-32, ADPCM tables, and Rijndael. We parse the \texttt{signsrch.sig} file provided with the tool as a reference for mapping each algorithm to its respective constants.

\subsection{Models}

We select several high-performing, transformer-based embedding models from the MTEB leaderboard\footnote{\url{https://huggingface.co/spaces/mteb/leaderboard}} \citep{muennighoff2023mtebmassivetextembedding}, hosted on Hugging Face, for comparison. We test BinSeek-Embedding \citep{chen2026crossmodalretrievalmodelsstripped}, our base model Qwen3-Embedding-0.6B \citep{zhang2025qwen3embeddingadvancingtext} and its 8B variant, bge-small-en-v1.5 \citep{xiao2024cpackpackedresourcesgeneral}, bge-m3 \citep{chen2025m3embeddingmultilingualitymultifunctionalitymultigranularity}, embeddinggemma-300m \citep{vera2025embeddinggemmapowerfullightweighttext}, snowflake-arctic-embed-l-v2.0 \citep{yu2024arcticembed20multilingualretrieval}, SFR-Embedding-Mistral \citep{SFRAIResearch2024}, jina-embeddings-v2-base-code \citep{gunther2024jinaembeddings28192token}, and nomic-embed-code \citep{suresh2025cornstackhighqualitycontrastivedata}. Each model outputs a single embedding vector per input.

We serve models with vLLM on GPU for accelerated inference, although this is not required for reproducing this work.

To reduce memory usage and increase throughput for large-scale embedding, we also quantize our model to FP8 format and evaluate its performance alongside the unquantized version.

\subsection{Metrics}

 Given two vectors \textbf{A} and \textbf{B}, we compute their cosine similarity:

\[sim_{\cos}(\mathbf{A}, \mathbf{B}) = \frac{\mathbf{A} \cdot \mathbf{B}}{\|\mathbf{A}\| \|\mathbf{B}\|}\]

For each query, we compute cosine similarity against all functions in the search pool and rank the results. We use Mean Reciprocal Rank (MRR), Recall, and Average Precision (AP) to measure search performance.

For each set of queries, we construct a vector of result rankings \textit{Q} by finding the rank of the single, desired search result per query. We express these ranks individually as \textit{q}. Some metrics incorporate information about whether a desired result is above a specified cutoff rank, which we define as the positive integer \textit{k}. For a result at rank position \textit{i}, we write \(\mathbf{1}[i \text{ is relevant}]\) to indicate whether that position holds a relevant result (1 if yes, 0 otherwise).

Mean Reciprocal Rank (MRR) and MRR@k (for single desired result):

\begin{align*}
\text{MRR} = \frac{1}{|Q|} \sum_{q=1}^{|Q|} \frac{1}{\text{rank}_q}  \qquad &
\text{MRR@k} = \frac{1}{|Q|} \sum_{q=1}^{|Q|} \frac{1}{\text{rank}_q} \cdot \mathbf{1}[\text{rank}_q \leq k]
\end{align*}

Recall@k (for single desired result):

\[\text{Recall@k} = \frac{1}{|Q|} \sum_{q=1}^{|Q|} \mathbf{1}[\text{rank}_q \leq k]\]

For Signsrch experiments, multiple correct query responses may be present. We let \textit{N} denote the total number of results in the ranked list, and \textit{R} denote the total number of relevant results for a given query. For these experiments, we use Average Precision (AP) and Mean Average Precision (MAP), where precision at rank \textit{i} is defined as:

\[
    P(i) = \frac{\displaystyle\sum_{j=1}^{i} \mathbf{1}[j \text{ is relevant}]}{i}
\]

Average Precision (AP) and Mean Average Precision (MAP):

\begin{align*}
    \text{AP}  &= \frac{1}{R} \sum_{i=1}^{N} \mathbf{1}[i \text{ is relevant}] \cdot P(i)  &
    \text{MAP} &= \frac{1}{|Q|} \sum_{q=1}^{|Q|} \text{AP}_q
\end{align*}

\section{Experiments}

\subsection{Training}

To study the effectiveness of the contrastive learning method for association between source and decompiled, stripped code, we fine-tune an existing Qwen3-Embedding-0.6B model. We train against the InfoNCE \citep{DBLP:journals/corr/abs-1807-03748} Loss objective using in-batch negatives and define anchor representations as decompiled functions and the positive/negative representations as source functions.

To compute InfoNCE loss, given $x$ = anchor (query) sample, $x^+$ = positive sample, $x^-_k$ = $k$-th negative sample, $K$ = number of negatives, \(sim_{\cos}\) = similarity function (we use cosine, see above), \(\tau\) = temperature:

\[\mathcal{L}_{\text{InfoNCE}} = -\log \frac{\exp(sim_{\cos}(x, x^+) / \tau)}{\exp(sim_{\cos}(x, x^+) / \tau) + \sum_{k=1}^{K} \exp(sim_{\cos}(x, x^-_k) / \tau)}\]

We set \(\tau = 0.05\) and use AdamW as our optimizer. We fine-tune Qwen3-Embedding-0.6B with an effective batch size of 512 over a single epoch of 964 steps with an initial learning rate of $2 \times 10^{-5}$, a cosine learning rate schedule, and a warmup ratio of 0.1. We train with bf16 mixed precision and hold out 1\% of the training data for validation. For training hardware, we use 4 NVIDIA H100s on an internal cluster. We refer to our model as \textit{BUSTA-IFCE-Qwen3-0.6B}.

\subsection{Experimental Setup}

\subsubsection{Assemblage Experiments}

We formalize two search scenarios over two distinct search pools. We assume a function is functionally equivalent in both its source and decompiled, stripped representations, allowing us to create a 1-to-1 relevant mapping between them.

\textbf{Source \textrightarrow \space Decompiled}: The analyst queries with source code to find the corresponding decompiled, stripped function.

\textbf{Decompiled \textrightarrow \space Source}: The analyst queries with decompiled, stripped code to find the corresponding source function.

\textbf{Filtered Search Pool} (Source Candidates Only, Decompiled Candidates Only): The search pool contains only the target representation. For example, if the analyst queries with decompiled code to find source code, the pool contains only source functions.

\textbf{Both Representations Search Pool} (Source and Decompiled Candidates): The search pool contains both source and decompiled, stripped functions. This makes the task harder because decompiled functions share similar generic variable and function names (e.g., \texttt{Var1}, \texttt{param\_1}), even though they may be functionally different. Models that rely too heavily on identifier names may match on naming similarity rather than functional equivalence.

Using two search pools helps characterize a model's reliance on these identifier names: only one functionally equivalent function exists in the pool, but many decompiled functions share similar or identical variable names due to naming conventions like \texttt{Var1}, \texttt{Var2}, \texttt{param\_1}, \texttt{param\_2}. As \citet{semanticcodeclones} demonstrate, certain code embedding models have been shown to rely heavily on identifier names, which may hinder performance at this task.

\subsubsection{Signsrch Experiments}

Signsrch data consists of algorithm names and corresponding constant values. To evaluate the extent to which models can associate these values, we measure performance on two tasks. Both tasks involve a single search pool of all available constants in the Signsrch dataset.

\textbf{Algorithm Group Name \textrightarrow \space Constants}: The analyst queries with an algorithm name to identify its relevant constants, requiring the model to produce meaningful natural language embeddings.

\textbf{Constant \textrightarrow \space Other Group Constants}: The analyst queries with a constant to find other constants in the same algorithm group, relying less on natural language understanding. The query constant is excluded from the search pool.

\subsubsection{Quantization}

We test inference with five trials on embedding 10,000 function pairs with a batch size of 64 on a single H100. This allows us to study the effect quantization has on memory requirements and model throughput. Reported memory usage reflects model weight loading only (from vLLM logs) and does not include batch encoding overhead.

\subsection{Experimental Results}

\subsubsection{Assemblage Experiments}

Results for \textbf{Decompiled \textrightarrow \space Source} are shown in Tables 1 and 2, and \textbf{Source \textrightarrow \space Decompiled} in Tables 3 and 4.

The combined search pools (Tables 1 and 3, Figures 1 and 2) contain 58,999 functions, while the filtered pools (Tables 2 and 4) contain 29,499.

Figure 1 plots Recall@k across all values of k for \textbf{Decompiled \textrightarrow \space Source} with a search pool of both representations. This shows approximately how deep into the ranked results an analyst must search to find the relevant function. Note that many models struggle with the first half of the combined search pool, suggesting that function representation type significantly affects model embedding.

Our model outperforms all others on every metric across all Assemblage search tasks and pool configurations.

\begin{figure}[htbp]
    \centering
    \includegraphics[width=0.8\textwidth]{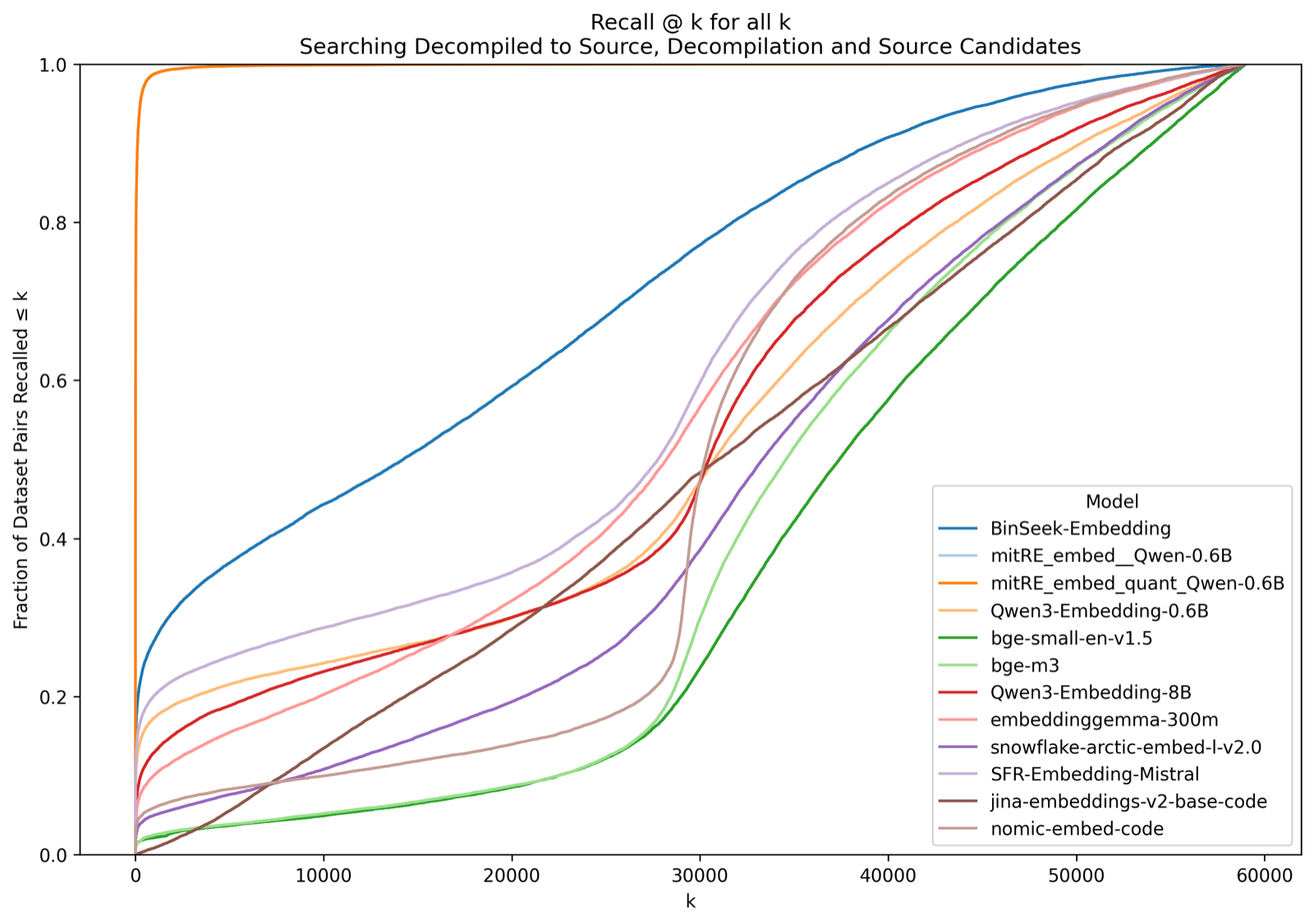}
    \caption{Recall@k, for all k, \textbf{Decompiled \textrightarrow \space Source}, Both Representations in Search Pool. Our model outperforms others and BinSeek-Embedding does not have the same difficulties other models have on the first half of the dataset.}
    \label{fig:label}
\end{figure}
\begin{figure}[htbp]
    \centering
    \includegraphics[width=0.9\textwidth]{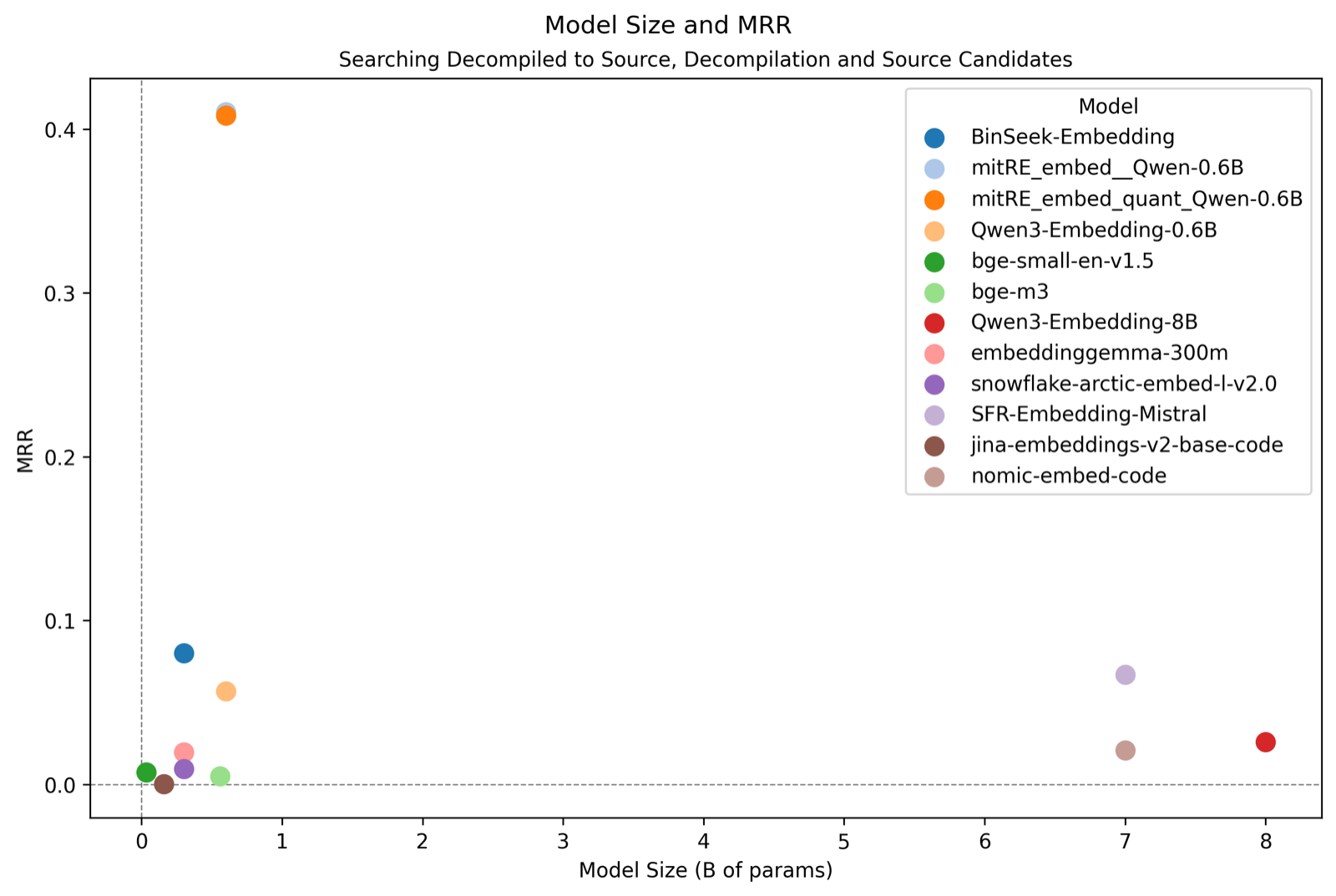}
    \caption{Model Size and MRR, \textbf{Decompiled \textrightarrow \space Source}, Both Representations in Search Pool. Model size appears unrelated to performance in this task.}
    \label{fig:label}
\end{figure}

\begin{table}
\footnotesize
\centering
\caption{MRR, MRR@k and Recall@k for Searching Decompiled to Source, Decompilation and Source Candidates.}
\label{tab:metrics_Searching_Decompiled_to_Source,_Decompilation_and_Source_Candidates}
\begin{tabular}{llrrrrrrrrr}
\toprule
Model & MRR & Recall@1 & Recall@5 & Recall@10 & MRR@10 \\
\midrule
BinSeek-Embedding & 0.0802 & 0.0594 & 0.1005 & 0.1221 & 0.0771 \\
\textbf{BUSTA-IFCE-Qwen3-0.6B} & \textbf{0.4104} & \textbf{0.3065} & \textbf{0.5176} & \textbf{0.6427} & \textbf{0.3983} \\
BUSTA-IFCE-Qwen3-0.6B-FP8 & 0.4083 & 0.3047 & 0.5162 & 0.6406 & 0.3961 \\
Qwen3-Embedding-0.6B & 0.0569 & 0.0434 & 0.0714 & 0.0822 & 0.0551 \\
bge-small-en-v1.5 & 0.0075 & 0.0056 & 0.0095 & 0.0113 & 0.0072 \\
bge-m3 & 0.0049 & 0.0034 & 0.0062 & 0.0078 & 0.0046 \\
Qwen3-Embedding-8B & 0.0258 & 0.0167 & 0.0346 & 0.0433 & 0.0242 \\
embeddinggemma-300m & 0.0196 & 0.0127 & 0.0259 & 0.0327 & 0.0184 \\
snowflake-arctic-embed-l-v2.0 & 0.0096 & 0.0061 & 0.0126 & 0.0162 & 0.0089 \\
SFR-Embedding-Mistral & 0.0671 & 0.0508 & 0.0842 & 0.0980 & 0.0649 \\
jina-embeddings-v2-base-code & 0.0001 & 0.0000 & 0.0000 & 0.0000 & 0.0000 \\
nomic-embed-code & 0.0209 & 0.0166 & 0.0253 & 0.0290 & 0.0203 \\
\bottomrule
\end{tabular}
\end{table}

\begin{table}
\footnotesize
\centering
\caption{MRR, MRR@k and Recall@k for Searching Decompiled to Source, Source Candidates Only.}
\label{tab:metrics_Searching_Decompiled_to_Source,_Source_Candidates_Only}
\begin{tabular}{llrrrrrrrrr}
\toprule
Model & MRR & Recall@1 & Recall@5 & Recall@10 & MRR@10 \\
\midrule
BinSeek-Embedding & 0.1656 & 0.1302 & 0.1998 & 0.2330 & 0.1603 \\
\textbf{BUSTA-IFCE-Qwen3-0.6B} & \textbf{0.6207} & \textbf{0.5112} & \textbf{0.7592} & \textbf{0.8353} & \textbf{0.6153} \\
BUSTA-IFCE-Qwen3-0.6B-FP8 & 0.6196 & 0.5097 & 0.7589 & 0.8349 & 0.6142 \\
Qwen3-Embedding-0.6B & 0.1526 & 0.1224 & 0.1836 & 0.2095 & 0.1486 \\
bge-small-en-v1.5 & 0.0391 & 0.0296 & 0.0472 & 0.0554 & 0.0374 \\
bge-m3 & 0.0633 & 0.0475 & 0.0776 & 0.0931 & 0.0606 \\
Qwen3-Embedding-8B & 0.1531 & 0.1126 & 0.1948 & 0.2290 & 0.1482 \\
embeddinggemma-300m & 0.1074 & 0.0793 & 0.1331 & 0.1593 & 0.1026 \\
snowflake-arctic-embed-l-v2.0 & 0.0673 & 0.0520 & 0.0815 & 0.0949 & 0.0646 \\
SFR-Embedding-Mistral & 0.2025 & 0.1634 & 0.2416 & 0.2752 & 0.1976 \\
jina-embeddings-v2-base-code & 0.0003 & 0.0000 & 0.0001 & 0.0003 & 0.0001 \\
nomic-embed-code & 0.1631 & 0.1214 & 0.2081 & 0.2452 & 0.1582 \\
\bottomrule
\end{tabular}
\end{table}

\begin{table}
\footnotesize
\centering
\caption{MRR, MRR@k and Recall@k for Searching Source to Decompiled, Decompilation and Source Candidates.}
\label{tab:metrics_Searching_Source_to_Decompiled,_Decompilation_and_Source_Candidates}
\begin{tabular}{llrrrrrrrrr}
\toprule
Model & MRR & Recall@1 & Recall@5 & Recall@10 & MRR@10 \\
\midrule
BinSeek-Embedding & 0.0815 & 0.0566 & 0.1029 & 0.1272 & 0.0768 \\
\textbf{BUSTA-IFCE-Qwen3-0.6B} & \textbf{0.4104} & \textbf{0.3090} & \textbf{0.5118} & \textbf{0.6337} & \textbf{0.3975} \\
BUSTA-IFCE-Qwen3-0.6B-FP8 & 0.4093 & 0.3082 & 0.5101 & 0.6308 & 0.3963 \\
Qwen3-Embedding-0.6B & 0.0994 & 0.0691 & 0.1285 & 0.1559 & 0.0944 \\
bge-small-en-v1.5 & 0.0251 & 0.0153 & 0.0332 & 0.0426 & 0.0230 \\
bge-m3 & 0.0236 & 0.0132 & 0.0318 & 0.0442 & 0.0212 \\
Qwen3-Embedding-8B & 0.0797 & 0.0485 & 0.1087 & 0.1394 & 0.0739 \\
embeddinggemma-300m & 0.0377 & 0.0216 & 0.0512 & 0.0683 & 0.0344 \\
snowflake-arctic-embed-l-v2.0 & 0.0501 & 0.0304 & 0.0673 & 0.0871 & 0.0466 \\
SFR-Embedding-Mistral & 0.1429 & 0.1028 & 0.1809 & 0.2180 & 0.1366 \\
jina-embeddings-v2-base-code & 0.0002 & 0.0000 & 0.0001 & 0.0003 & 0.0001 \\
nomic-embed-code & 0.0760 & 0.0475 & 0.1013 & 0.1287 & 0.0705 \\
\bottomrule
\end{tabular}
\end{table}

\begin{table}
\footnotesize
\centering
\caption{MRR, MRR@k and Recall@k for Searching Source to Decompiled, Decompilation Candidates Only.}
\label{tab:metrics_Searching_Source_to_Decompiled,_Decompilation_Candidates_Only}
\begin{tabular}{llrrrrrrrrr}
\toprule
Model & MRR & Recall@1 & Recall@5 & Recall@10 & MRR@10 \\
\midrule
BinSeek-Embedding & 0.1618 & 0.1238 & 0.1993 & 0.2345 & 0.1562 \\
\textbf{BUSTA-IFCE-Qwen3-0.6B} & \textbf{0.5962} & \textbf{0.4881} & \textbf{0.7292} & \textbf{0.8094} & \textbf{0.5898} \\
BUSTA-IFCE-Qwen3-0.6B-FP8 & 0.5950 & 0.4867 & 0.7285 & 0.8085 & 0.5886 \\
Qwen3-Embedding-0.6B & 0.1821 & 0.1459 & 0.2195 & 0.2501 & 0.1777 \\
bge-small-en-v1.5 & 0.0701 & 0.0547 & 0.0842 & 0.0983 & 0.0676 \\
bge-m3 & 0.0973 & 0.0753 & 0.1187 & 0.1392 & 0.0939 \\
Qwen3-Embedding-8B & 0.2035 & 0.1597 & 0.2486 & 0.2864 & 0.1978 \\
embeddinggemma-300m & 0.1461 & 0.1121 & 0.1800 & 0.2101 & 0.1413 \\
snowflake-arctic-embed-l-v2.0 & 0.1352 & 0.1078 & 0.1630 & 0.1861 & 0.1317 \\
SFR-Embedding-Mistral & 0.2294 & 0.1862 & 0.2742 & 0.3107 & 0.2242 \\
jina-embeddings-v2-base-code & 0.0003 & 0.0000 & 0.0002 & 0.0004 & 0.0001 \\
nomic-embed-code & 0.2208 & 0.1756 & 0.2672 & 0.3061 & 0.2152 \\
\bottomrule
\end{tabular}
\end{table}

\subsubsection{Signsrch Experiments}

We report results for the Signsrch Experiments in Table 5. Note that because each group may contain a different number of constants, for the \textbf{Constant \textrightarrow \space Other Constants} experiment we compute MAP within each group and then average across groups (macro-averaged MAP). This ensures equal weighting across groups regardless of size.

\begin{table}[ht]
  \caption{Signsrch Experiment Result Metric Means and Standard Errors. \textit{Embeddinggemma-300m} excels at \textbf{Group Name \textrightarrow \space Constants} and our model excels at \textbf{Constant \textrightarrow \space Other Constants}. Standard error of the mean is denoted as S.e.}

  \label{tab:ap_statistics}
    \scriptsize
  \centering
  \begin{tabular}{l@{\hspace{6pt}}c@{\hspace{4pt}}c@{\hspace{8pt}}c@{\hspace{4pt}}c}
    Model & \multicolumn{2}{c}{\textbf{Group \textrightarrow \space Constants}} & \multicolumn{2}{c}{\textbf{Constant \textrightarrow \space Other Constants}} \\ \hline
    \hline
    \space & \multicolumn{2}{c}{Avg. Precision} & \multicolumn{2}{c}{Mean Avg. Precision} \\ \hline
    \space & Mean & S.e. & Mean & S.e. \\
    \hline
    BinSeek-Embedding & 0.018303 & 0.003423 & 0.037864 & 0.002213 \\
    \textbf{BUSTA-IFCE-Qwen3-0.6B} & 0.019732 & 0.003391 & \textbf{0.062765} & 0.002534 \\
    BUSTA-IFCE-Qwen3-0.6B-FP8 & 0.020190 & 0.003461 & 0.062490 & 0.002562  \\
    Qwen3-Embedding-0.6B & 0.019477 & 0.003444 & 0.038412 & 0.002413 \\
    bge-small-en-v1.5 & 0.014982 & 0.003174 & 0.018950 & 0.001430  \\
    bge-m3 & 0.018454 & 0.003532 & 0.024086 & 0.001626\\
    Qwen3-Embedding-8B & 0.019321 & 0.003234 &  0.024086 & 0.001626\\
    \textbf{embeddinggemma-300m} & \textbf{0.022182} & 0.003545 & 0.027630 & 0.001650 \\
    snowflake-arctic-embed-l-v2.0 & 0.017510 & 0.003427 & 0.034196 & 0.002067\\
    SFR-Embedding-Mistral & 0.015586 & 0.002970 & 0.025607 & 0.001750 \\
    jina-embeddings-v2-base-code & 0.002623 & 0.000235 & 0.002767 & 0.000229\\
    nomic-embed-code & 0.019272 & 0.003469 & 0.002967 & 0.015254 \\ 
    \hline
  \end{tabular}
\end{table}

Shapiro-Wilk testing reveals that all distributions of model Average Precision and Mean Average Precision scores are not normally distributed. To rigorously compare model performance, we conduct pairwise one-sided Wilcoxon tests (matched, nonparametric). Consistent with the mean values in Table 5, \textit{embeddinggemma-300m} outperforms all others at the task of \textbf{Algorithm Group Name \textrightarrow \space Constants} and our model outperforms all others at \textbf{Constant \textrightarrow \space Other Group Constants}. However, our model does not significantly outperform others at \textbf{Algorithm Group Name \textrightarrow \space Constants}, placing mid-range among all models tested.

\subsubsection{Quantization}

Across five trials embedding 10,000 function-pair instances, the FP8-quantized model required, on average, 88.6\% of the inference time of the full-precision model (standard deviation = 1.16\%). The quantized model also uses approximately 64\% of the memory, consuming 0.72 GB vs. 1.12 GB of VRAM for model weights.

\section{Discussion}

\paragraph{Contrastive Learning can teach models to embed functions based on functionality} In every Assemblage experiment, our model outperforms others by a large margin. This demonstrates that contrastive learning enables embedding models to associate functions across representations, even without informative identifier names. This improvement could serve as the basis for production search tools that help binary reverse engineers identify functionally equivalent code across source and decompiled representations.

Performance drops on the combined search pools suggest other models rely heavily on identifier names. The recall curve in Figure 1 illustrates this: many models struggle to recall the correct function for the first half of the search pool, followed by a steep increase in recall near the midpoint. This confirms our expectation that models may struggle to discriminate well between decompiled functions, even when the composition of the function is very different.

Despite not being trained for this task, BinSeek-Embedding exhibits this behavior to a lesser degree than other open-source models on the combined search pool. This suggests that training a model to associate natural language with decompiled functions may implicitly improve discrimination between decompiled functions, a desirable but previously unreported effect.

\paragraph{Signsrch experiments demonstrate model generalization to a new task} Our model was never explicitly trained to perform constant association, yet achieves improvement over its base model and others at the \textbf{Constant \textrightarrow \space Other Group Constants} task. This may be attributed to an overlap of constants within the Assemblage dataset relevant to the constants in Signsrch. This supports the idea that our model learns meaningful features and has not simply overfit to Assemblage's data format and composition.

\paragraph{Quantization does not entirely hinder model performance}
On most benchmarks, our quantized model (FP8) performs comparably to the unquantized version. This suggests the approach is viable in low-compute environments where conserving VRAM enables larger batch sizes and higher throughput.

\section{Conclusion}
We present a framework for evaluating how well embedding models can associate source code functions with their decompiled, stripped counterparts that lack identifiers like variable, parameter, and function names. We demonstrate that contrastive learning on a paired corpus such as Assemblage can substantially improve function association, achieving approximately 5x improvement in MRR over the next best model on combined search pools during \textbf{Decompiled \textrightarrow \space Source} association. Our results show that the combined search pools, which include both representations, present significantly harder tasks for models that rely on identifier names, and that our model reduces this reliance. Finally, we show that our model generalizes to a separate constant-association task it was never trained on, indicating that it learns broader algorithmic features not dependent on identifier names.

\section{Limitations}

These results may be limited by the underlying data distribution of the Assemblage dataset: compilation only for x86 means that model performance may deviate on binaries compiled for different architectures. Since the Assemblage dataset also only has labeled source code for its WinPE split, our model and evaluations only reflect the distribution of functions found and compiled automatically from the dataset authors' web crawl, which almost certainly does not encompass all of the software a binary reverse engineer might be interested in. Also, vulnerability analysis is an important motivation for many binary reverse engineers, yet there are no guarantees that this data contains interesting or meaningful vulnerabilities. Our evaluations also do not compare models to the existing, rules-based tools mentioned in the Introduction.

Under a similar consideration, Signsrch only encompasses a finite set of constants that may not represent the full breadth of interesting reverse engineering data. We seek an equal weighting of each algorithm group in our evaluation, but it is possible there is an imbalance of group categories themselves (i.e., more cryptography than multimedia).

\section{Impacts}

The intent of this work is to empower systems that might help binary reverse engineers perform defensive activities. However, such systems may also be used for malicious purposes. For example, distributing these methods could lead an attacker to use models to map an existing vulnerability (present in a library, for example) to a given binary and use this for exploitation. This harm is an example of an intentional misuse of our technology. To mitigate such harm, we recommend future works to refrain from using high-sensitivity data for models and evaluations.

\begin{ack}

Thank you to Peter Lucia, Corre Steele, Dr. Chris Niessen, and Marissa Dotter for their helpful feedback during peer review. Thank you to Dr. Stan Barr for guiding this work, thoroughly reviewing it, and supporting funding acquisition. Thank you to Deirdre Murphy for helping with the Assemblage dataset.

This research was funded by MITRE’s Independent Research and Development Program.
\end{ack}

\bibliography{refs}

@article{Wang_Zhou_Han_Zhang_2026, title={BDLF-Qwen3: Enhanced Cross-Architecture Binary Function Similarity Detection Through Binary Dynamic Layer Fusion}, volume={40}, url={https://ojs.aaai.org/index.php/AAAI/article/view/37094}, DOI={10.1609/aaai.v40i2.37094}, abstractNote={Binary code analysis is essential for software security across various instruction set architectures. Cross-architecture binary function similarity detection faces significant challenges due to substantial differences in instruction sets and architectural conventions. Existing approaches struggle to capture relationships between code abstraction levels, and lack comprehensive cross-architecture datasets for effective evaluation. Inspired by human cognitive processes of dynamically integrating multi-level information, we propose Binary Dynamic Layer Fusion (BDLF), a novel neural architecture that enhances cross-architecture similarity detection through adaptive layer-wise feature integration. BDLF leverages Qwen3’s multilingual code understanding and introduces dynamic weight generation to optimally combine representations from all previous layers. We also construct Cross-Bin, a high quality cross-architecture binary function dataset. BDLF-Qwen3 employs two-stage training: partial fine-tuning with pairwise similarity learning followed by BDLF enhancement with InfoNCE contrastive learning. Experiments demonstrate BDLF-Qwen3 significantly outperforms state-of-the-art methods, achieving 36-65\% improvement in Recall@10 across diverse CPU architectures.}, number={2}, journal={Proceedings of the AAAI Conference on Artificial Intelligence}, author={Wang, Yuanda and Zhou, Ji and Han, Xinhui and Zhang, Chao}, year={2026}, month={Mar.}, pages={1222-1230} }

@misc{chen2026crossmodalretrievalmodelsstripped,
      title={Cross-modal Retrieval Models for Stripped Binary Analysis}, 
      author={Guoqiang Chen and Lingyun Ying and Ziyang Song and Daguang Liu and Qiang Wang and Zhiqi Wang and Li Hu and Shaoyin Cheng and Weiming Zhang and Nenghai Yu},
      year={2026},
      eprint={2512.10393},
      archivePrefix={arXiv},
      primaryClass={cs.SE},
      url={https://arxiv.org/abs/2512.10393}, 
}

@article{YU2025129238,
title = {CrossCode2Vec: A unified representation across source and binary functions for code similarity detection},
journal = {Neurocomputing},
volume = {620},
pages = {129238},
year = {2025},
issn = {0925-2312},
doi = {https://doi.org/10.1016/j.neucom.2024.129238},
url = {https://www.sciencedirect.com/science/article/pii/S0925231224020095},
author = {Gaoqing Yu and Jing An and Jiuyang Lyu and Wei Huang and Wenqing Fan and Yixuan Cheng and Aina Sui}
}

@misc{zhang2025pretrainingrepresentationsbinarycode,
      title={Pre-Training Representations of Binary Code Using Contrastive Learning}, 
      author={Yifan Zhang and Chen Huang and Yueke Zhang and Huajie Shao and Kevin Leach and Yu Huang},
      year={2025},
      eprint={2210.05102},
      archivePrefix={arXiv},
      primaryClass={cs.SE},
      url={https://arxiv.org/abs/2210.05102}, 
}

@article{Jia2024SemanticAwareInstructionEmbedding,
  author  = {Jia, Y. and Yu, Z. and Hong, Z.},
  title   = {Semantic aware-based instruction embedding for binary code similarity detection},
  journal = {PLOS ONE},
  year    = {2024},
  month   = jun,
  volume  = {19},
  number  = {6},
  pages   = {e0305299},
  doi     = {10.1371/journal.pone.0305299},
  pmid    = {38861533},
  pmcid   = {PMC11166306}
}

@article{10.1145/3728927,
author = {Zhang, Bolun and Gao, Zeyu and Wang, Hao and Cui, Yuxin and Qin, Siliang and Zhang, Chao and Chen, Kai and Zhao, Beibei},
title = {BinQuery: A Novel Framework for Natural Language-Based Binary Code Retrieval},
year = {2025},
issue_date = {July 2025},
publisher = {Association for Computing Machinery},
address = {New York, NY, USA},
volume = {2},
number = {ISSTA},
url = {https://doi.org/10.1145/3728927},
doi = {10.1145/3728927},
journal = {Proc. ACM Softw. Eng.},
month = jun,
articleno = {ISSTA052},
numpages = {23}
}

@inproceedings{Li_2021, series={CCS ’21},
   title={PalmTree: Learning an Assembly Language Model for Instruction Embedding},
   url={http://dx.doi.org/10.1145/3460120.3484587},
   DOI={10.1145/3460120.3484587},
   booktitle={Proceedings of the 2021 ACM SIGSAC Conference on Computer and Communications Security},
   publisher={ACM},
   author={Li, Xuezixiang and Qu, Yu and Yin, Heng},
   year={2021},
   month=nov, pages={3236–3251},
   collection={CCS ’21} }

@inproceedings{liu2024assemblageautomaticbinarydataset,
 author = {Liu, Chang and Saul, Rebecca and Sun, Yihao and Raff, Edward and Fuchs, Maya and Southard Pantano, Townsend and Holt, James and Micinski, Kristopher},
 booktitle = {Advances in Neural Information Processing Systems},
 doi = {10.52202/079017-1871},
 editor = {A. Globerson and L. Mackey and D. Belgrave and A. Fan and U. Paquet and J. Tomczak and C. Zhang},
 pages = {58698--58715},
 publisher = {Curran Associates, Inc.},
 title = {Assemblage: Automatic Binary Dataset Construction for Machine Learning},
 url = {https://proceedings.neurips.cc/paper_files/paper/2024/file/6bbefc73a187dd42e0dc065b4e7a0615-Paper-Datasets_and_Benchmarks_Track.pdf},
 volume = {37},
 year = {2024}
}

@inproceedings{NEURIPS2020_285f89b8,
 author = {Yu, Zeping and Zheng, Wenxin and Wang, Jiaqi and Tang, Qiyi and Nie, Sen and Wu, Shi},
 booktitle = {Advances in Neural Information Processing Systems},
 editor = {H. Larochelle and M. Ranzato and R. Hadsell and M.F. Balcan and H. Lin},
 pages = {3872--3883},
 publisher = {Curran Associates, Inc.},
 title = {CodeCMR: Cross-Modal Retrieval For Function-Level Binary Source Code Matching},
 url = {https://proceedings.neurips.cc/paper_files/paper/2020/file/285f89b802bcb2651801455c86d78f2a-Paper.pdf},
 volume = {33},
 year = {2020}
}

@misc{zhang2025qwen3embeddingadvancingtext,
      title={Qwen3 Embedding: Advancing Text Embedding and Reranking Through Foundation Models}, 
      author={Yanzhao Zhang and Mingxin Li and Dingkun Long and Xin Zhang and Huan Lin and Baosong Yang and Pengjun Xie and An Yang and Dayiheng Liu and Junyang Lin and Fei Huang and Jingren Zhou},
      year={2025},
      eprint={2506.05176},
      archivePrefix={arXiv},
      primaryClass={cs.CL},
      url={https://arxiv.org/abs/2506.05176}, 
}

@inproceedings{chen2025m3embeddingmultilingualitymultifunctionalitymultigranularity,
    title = "{M}3-Embedding: Multi-Linguality, Multi-Functionality, Multi-Granularity Text Embeddings Through Self-Knowledge Distillation",
    author = "Chen, Jianlyu  and
      Xiao, Shitao  and
      Zhang, Peitian  and
      Luo, Kun  and
      Lian, Defu  and
      Liu, Zheng",
    editor = "Ku, Lun-Wei  and
      Martins, Andre  and
      Srikumar, Vivek",
    booktitle = "Findings of the Association for Computational Linguistics: ACL 2024",
    month = aug,
    year = "2024",
    address = "Bangkok, Thailand",
    publisher = "Association for Computational Linguistics",
    url = "https://aclanthology.org/2024.findings-acl.137/",
    doi = "10.18653/v1/2024.findings-acl.137",
    pages = "2318--2335"
}

@inproceedings{xiao2024cpackpackedresourcesgeneral,
author = {Xiao, Shitao and Liu, Zheng and Zhang, Peitian and Muennighoff, Niklas and Lian, Defu and Nie, Jian-Yun},
title = {C-Pack: Packed Resources For General Chinese Embeddings},
year = {2024},
isbn = {9798400704314},
publisher = {Association for Computing Machinery},
address = {New York, NY, USA},
url = {https://doi.org/10.1145/3626772.3657878},
doi = {10.1145/3626772.3657878},
booktitle = {Proceedings of the 47th International ACM SIGIR Conference on Research and Development in Information Retrieval},
pages = {641–649},
numpages = {9},
location = {Washington DC, USA},
series = {SIGIR '24}
}

@misc{vera2025embeddinggemmapowerfullightweighttext,
      title={EmbeddingGemma: Powerful and Lightweight Text Representations}, 
      author={Henrique Schechter Vera and Sahil Dua and Biao Zhang and Daniel Salz and Ryan Mullins and Sindhu Raghuram Panyam and Sara Smoot and Iftekhar Naim and Joe Zou and Feiyang Chen and Daniel Cer and Alice Lisak and Min Choi and Lucas Gonzalez and Omar Sanseviero and Glenn Cameron and Ian Ballantyne and Kat Black and Kaifeng Chen and Weiyi Wang and Zhe Li and Gus Martins and Jinhyuk Lee and Mark Sherwood and Juyeong Ji and Renjie Wu and Jingxiao Zheng and Jyotinder Singh and Abheesht Sharma and Divyashree Sreepathihalli and Aashi Jain and Adham Elarabawy and AJ Co and Andreas Doumanoglou and Babak Samari and Ben Hora and Brian Potetz and Dahun Kim and Enrique Alfonseca and Fedor Moiseev and Feng Han and Frank Palma Gomez and Gustavo Hernández Ábrego and Hesen Zhang and Hui Hui and Jay Han and Karan Gill and Ke Chen and Koert Chen and Madhuri Shanbhogue and Michael Boratko and Paul Suganthan and Sai Meher Karthik Duddu and Sandeep Mariserla and Setareh Ariafar and Shanfeng Zhang and Shijie Zhang and Simon Baumgartner and Sonam Goenka and Steve Qiu and Tanmaya Dabral and Trevor Walker and Vikram Rao and Waleed Khawaja and Wenlei Zhou and Xiaoqi Ren and Ye Xia and Yichang Chen and Yi-Ting Chen and Zhe Dong and Zhongli Ding and Francesco Visin and Gaël Liu and Jiageng Zhang and Kathleen Kenealy and Michelle Casbon and Ravin Kumar and Thomas Mesnard and Zach Gleicher and Cormac Brick and Olivier Lacombe and Adam Roberts and Qin Yin and Yunhsuan Sung and Raphael Hoffmann and Tris Warkentin and Armand Joulin and Tom Duerig and Mojtaba Seyedhosseini},
      year={2025},
      eprint={2509.20354},
      archivePrefix={arXiv},
      primaryClass={cs.CL},
      url={https://arxiv.org/abs/2509.20354}, 
}

@misc{suresh2025cornstackhighqualitycontrastivedata,
      title={CoRNStack: High-Quality Contrastive Data for Better Code Retrieval and Reranking}, 
      author={Tarun Suresh and Revanth Gangi Reddy and Yifei Xu and Zach Nussbaum and Andriy Mulyar and Brandon Duderstadt and Heng Ji},
      year={2025},
      eprint={2412.01007},
      archivePrefix={arXiv},
      primaryClass={cs.CL},
      url={https://arxiv.org/abs/2412.01007}, 
}

@misc{yu2024arcticembed20multilingualretrieval,
      title={Arctic-Embed 2.0: Multilingual Retrieval Without Compromise}, 
      author={Puxuan Yu and Luke Merrick and Gaurav Nuti and Daniel Campos},
      year={2024},
      eprint={2412.04506},
      archivePrefix={arXiv},
      primaryClass={cs.CL},
      url={https://arxiv.org/abs/2412.04506}, 
}

@misc{SFRAIResearch2024,
  title={SFR-Embedding-Mistral: Enhance Text Retrieval with Transfer Learning},
  author={Rui Meng and Ye Liu and Shafiq Rayhan Joty and Caiming Xiong and Yingbo Zhou and Semih Yavuz},
  howpublished={Salesforce AI Research Blog},
  year={2024},
  url={https://www.salesforce.com/blog/sfr-embedding/}
}

@misc{gunther2024jinaembeddings28192token,
      title={Jina Embeddings 2: 8192-Token General-Purpose Text Embeddings for Long Documents}, 
      author={Michael Günther and Jackmin Ong and Isabelle Mohr and Alaeddine Abdessalem and Tanguy Abel and Mohammad Kalim Akram and Susana Guzman and Georgios Mastrapas and Saba Sturua and Bo Wang and Maximilian Werk and Nan Wang and Han Xiao},
      year={2024},
      eprint={2310.19923},
      archivePrefix={arXiv},
      primaryClass={cs.CL},
      url={https://arxiv.org/abs/2310.19923}, 
}

@inproceedings{muennighoff2023mtebmassivetextembedding,
    title = "{MTEB}: Massive Text Embedding Benchmark",
    author = "Muennighoff, Niklas  and
      Tazi, Nouamane  and
      Magne, Loic  and
      Reimers, Nils",
    editor = "Vlachos, Andreas  and
      Augenstein, Isabelle",
    booktitle = "Proceedings of the 17th Conference of the European Chapter of the Association for Computational Linguistics",
    month = may,
    year = "2023",
    address = "Dubrovnik, Croatia",
    publisher = "Association for Computational Linguistics",
    url = "https://aclanthology.org/2023.eacl-main.148/",
    doi = "10.18653/v1/2023.eacl-main.148",
    pages = "2014--2037"
}

@INPROCEEDINGS{word2vecmal,
  author={Popov, Igor},
  booktitle={2017 Siberian Symposium on Data Science and Engineering (SSDSE)}, 
  title={Malware detection using machine learning based on word2vec embeddings of machine code instructions}, 
  year={2017},
  volume={},
  number={},
  pages={1-4},
  doi={10.1109/SSDSE.2017.8071952}}

@article{semanticcodeclones,
author = {Yu, Hao and Hu, Xing and Li, Ge and Li, Ying and Wang, Qianxiang and Xie, Tao},
title = {Assessing and Improving an Evaluation Dataset for Detecting Semantic Code Clones via Deep Learning},
year = {2022},
issue_date = {October 2022},
publisher = {Association for Computing Machinery},
address = {New York, NY, USA},
volume = {31},
number = {4},
issn = {1049-331X},
url = {https://doi.org/10.1145/3502852},
doi = {10.1145/3502852},
journal = {ACM Trans. Softw. Eng. Methodol.},
month = jul,
articleno = {62},
numpages = {25}
}

@misc{signsrch,
year = {2016},
title={MyToolz},
author={Luigi Auriemma},
url = {https://aluigi.altervista.org/mytoolz.htm}}

@misc{flirt,
title={{FLIRT}: Fast Library Identification and Recognition Technology},
author={{Hex-Rays}},
year={2024},
url = {https://docs.hex-rays.com/user-guide/signatures/flirt},
note = {IDA Pro Documentation}
}

@misc{BSim,
title={{BSim}: Tutorial},
author={{National Security Agency}},
year={2026},
url = {https://ghidra.re/ghidra_docs/GhidraClass/BSim/README.html},
note = {BSim Documentation}
}

@misc{imgui,
  author = {{Cornut, Omar}},
  title = {{Dear ImGui}},
  year = {2024},
  url = {https://github.com/ocornut/imgui/tree/master},
}

@article{DBLP:journals/corr/abs-1807-03748,
  author       = {A{\"{a}}ron van den Oord and
                  Yazhe Li and
                  Oriol Vinyals},
  title        = {Representation Learning with Contrastive Predictive Coding},
  journal      = {CoRR},
  volume       = {abs/1807.03748},
  year         = {2018},
  url          = {http://arxiv.org/abs/1807.03748},
  eprinttype   = {arXiv},
  eprint       = {1807.03748},
  bibsource    = {dblp computer science bibliography, https://dblp.org}
}


\appendix

\section{Embedding Space Dimensionality Reduction}

We embed with different models, reduce in dimensionality with t-SNE, and plot 5000 function pairs from Assemblage colored by representation type. These plots, shown in Figures 3-6, are useful for exploratory data analysis. Separation in representation across embedding space may suggest (but does not prove) that for many open source models, a large amount of variance in embedding can be explained by the representation type and not by the operations actually conducted in the function itself (which should be identical across representations).

\begin{figure}[htbp]
    \centering
    \includegraphics[width=0.9\textwidth]{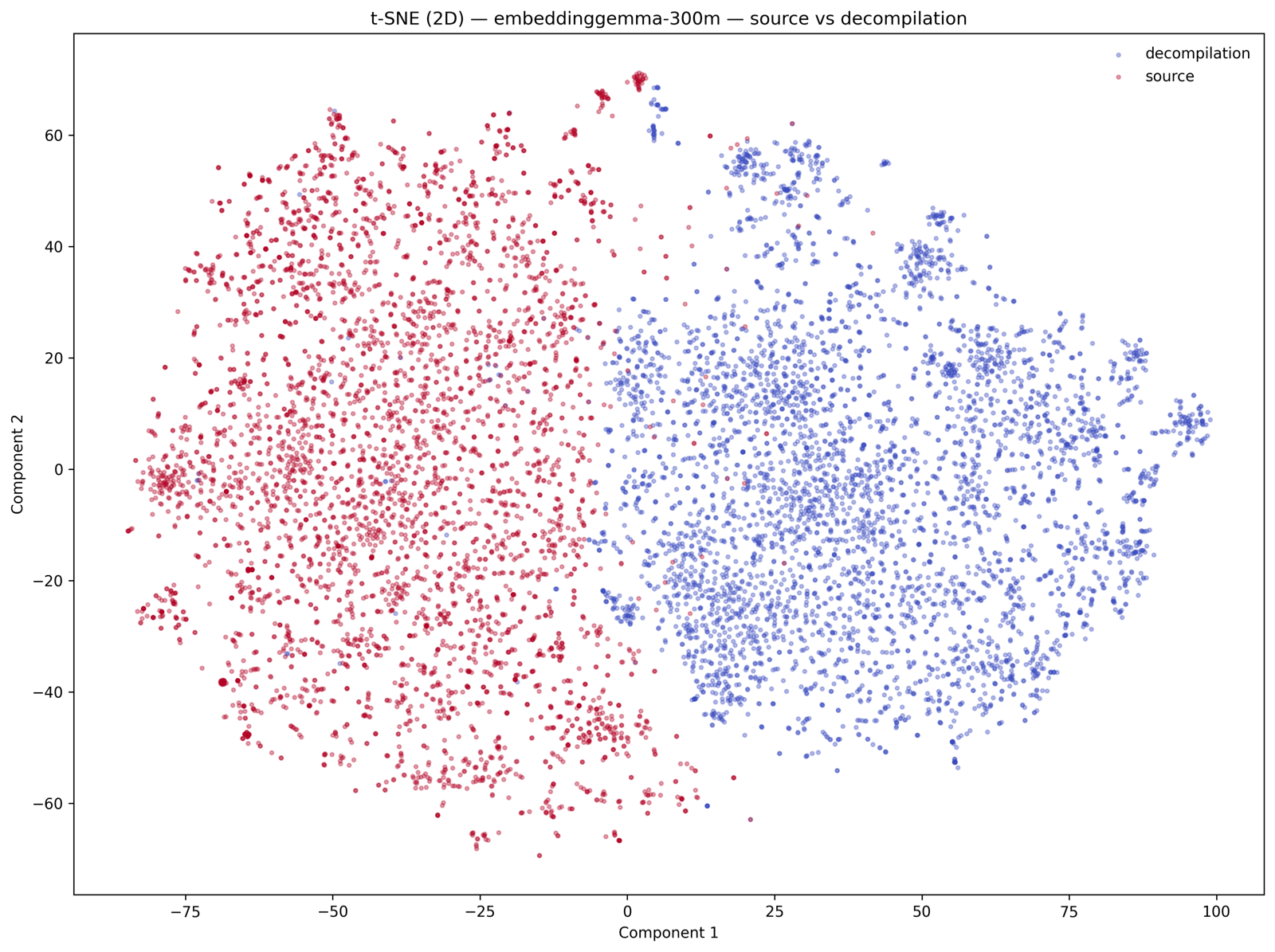}
    \caption{embeddinggemma-300 embedding space, reduced for visualization.}
    \label{fig:gemma-tsne}
\end{figure}

\begin{figure}[htbp]
    \centering
    \includegraphics[width=0.9\textwidth]{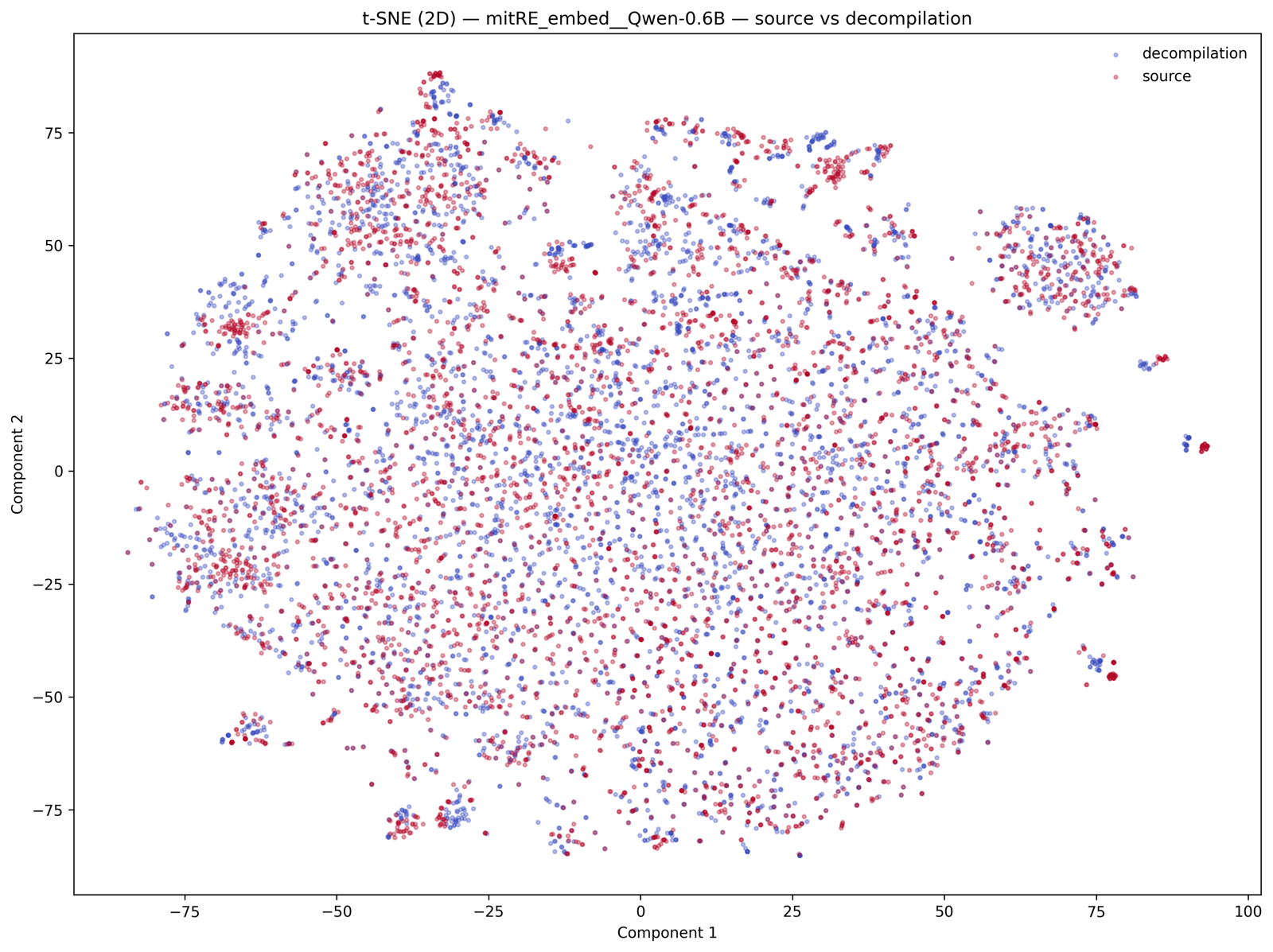}
    \caption{BUSTA-IFCE-Qwen3-0.6B embedding space, reduced for visualization.}
    \label{fig:our-tsne}
\end{figure}

\begin{figure}[htbp]
    \centering
    \includegraphics[width=0.9\textwidth]{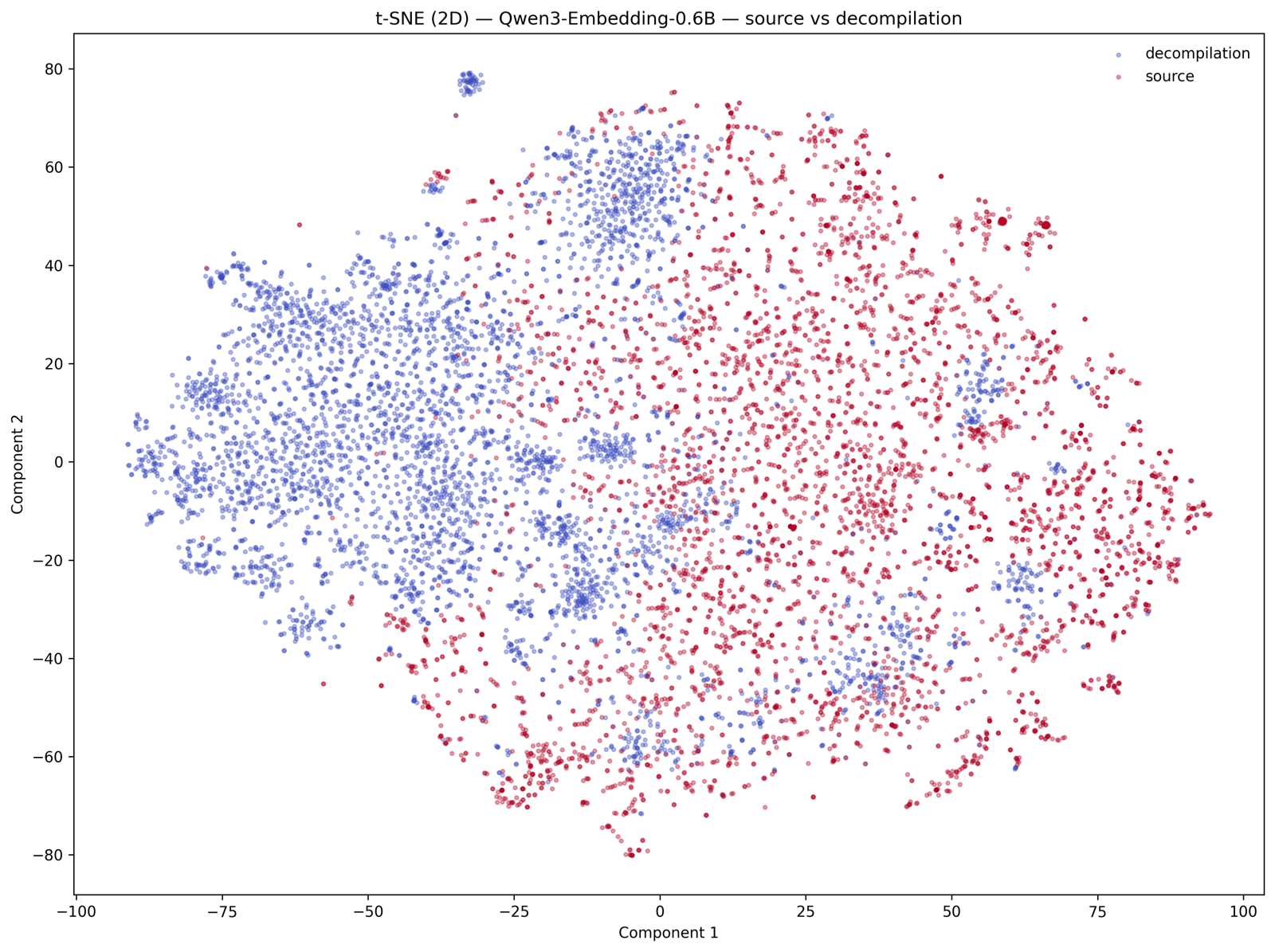}
    \caption{Qwen-3-Embedding-0.6B embedding space, reduced for visualization.}
    \label{fig:qwen-tsne}
\end{figure}

\begin{figure}[htbp]
    \centering
    \includegraphics[width=0.9\textwidth]{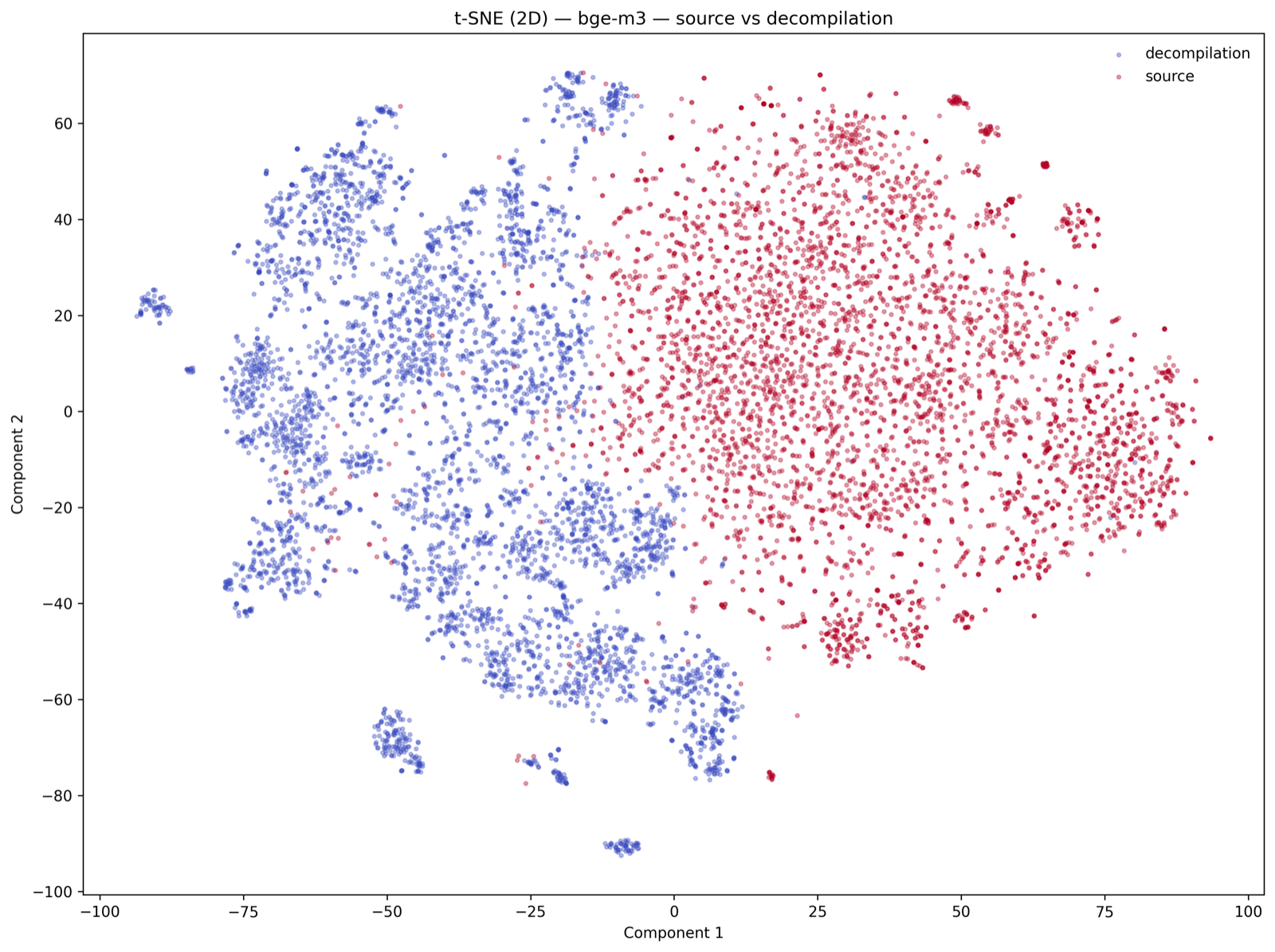}
    \caption{BGE-m3 embedding space, reduced for visualization.}
    \label{fig:bge-tsne}
\end{figure}

\section{Pairwise Wilcoxon Test Results}

In Figures 7 and 8, we visualize in a heatmap the results from pairwise Wilcoxon testing for each of the Signsrch experiments. Each value is a p-value from the one-sided statistical tests.

\begin{figure}[htbp]
    \centering
    \includegraphics[width=0.8\textwidth]{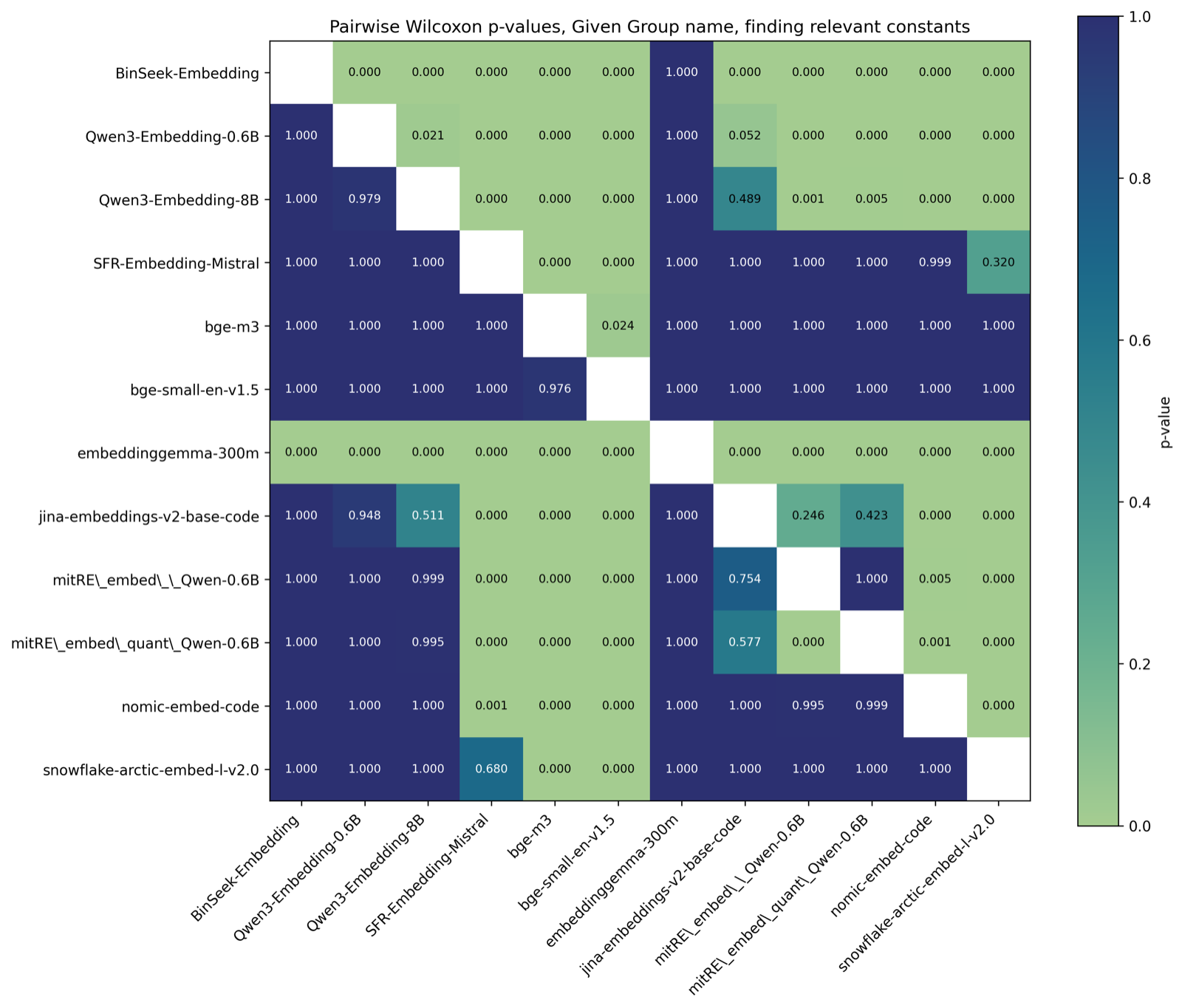}
    \caption{Signsrch results for \textbf{Algorithm Group Name \textrightarrow \space Constants}. Each value is the probability (p-value) that the model along the x axis had a greater score distribution than the model along the y. Embeddinggemma-300m exhibits the best performance.}
    \label{fig:bge-tsne}
\end{figure}

\begin{figure}[htbp]
    \centering
    \includegraphics[width=0.8\textwidth]{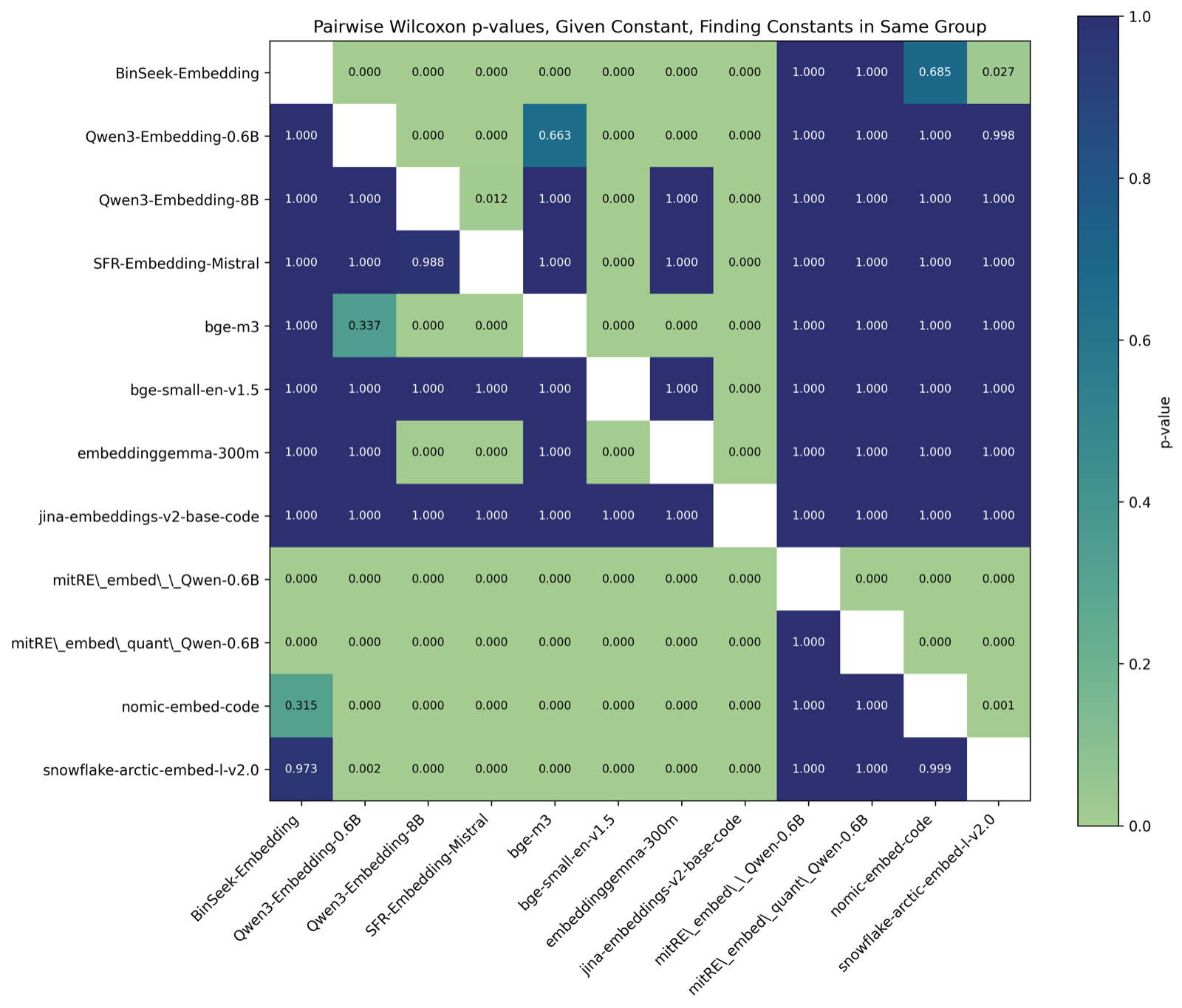}
    \caption{Signsrch results for \textbf{Constant \textrightarrow \space Other Group Constants}. Each value is the probability (p-value) that the model along the x axis had a greater score distribution than the model along the y. Our model exhibits the best performance.}
    \label{fig:bge-tsne}
\end{figure}

\section{Assemblage Experiment MRR Bar Graphs with Error}

To help visualize the results of our Assemblage experiments, we plot the MRR of each model with standard error bars in Figures 9-12. These standard error values are standard error of the mean and reflect variance across queries. Note the function count indicates the size of that dataset split, the actual search pool contains one less function because the query function is removed.

\begin{figure}[htbp]
    \centering
    \includegraphics[width=0.9\textwidth]{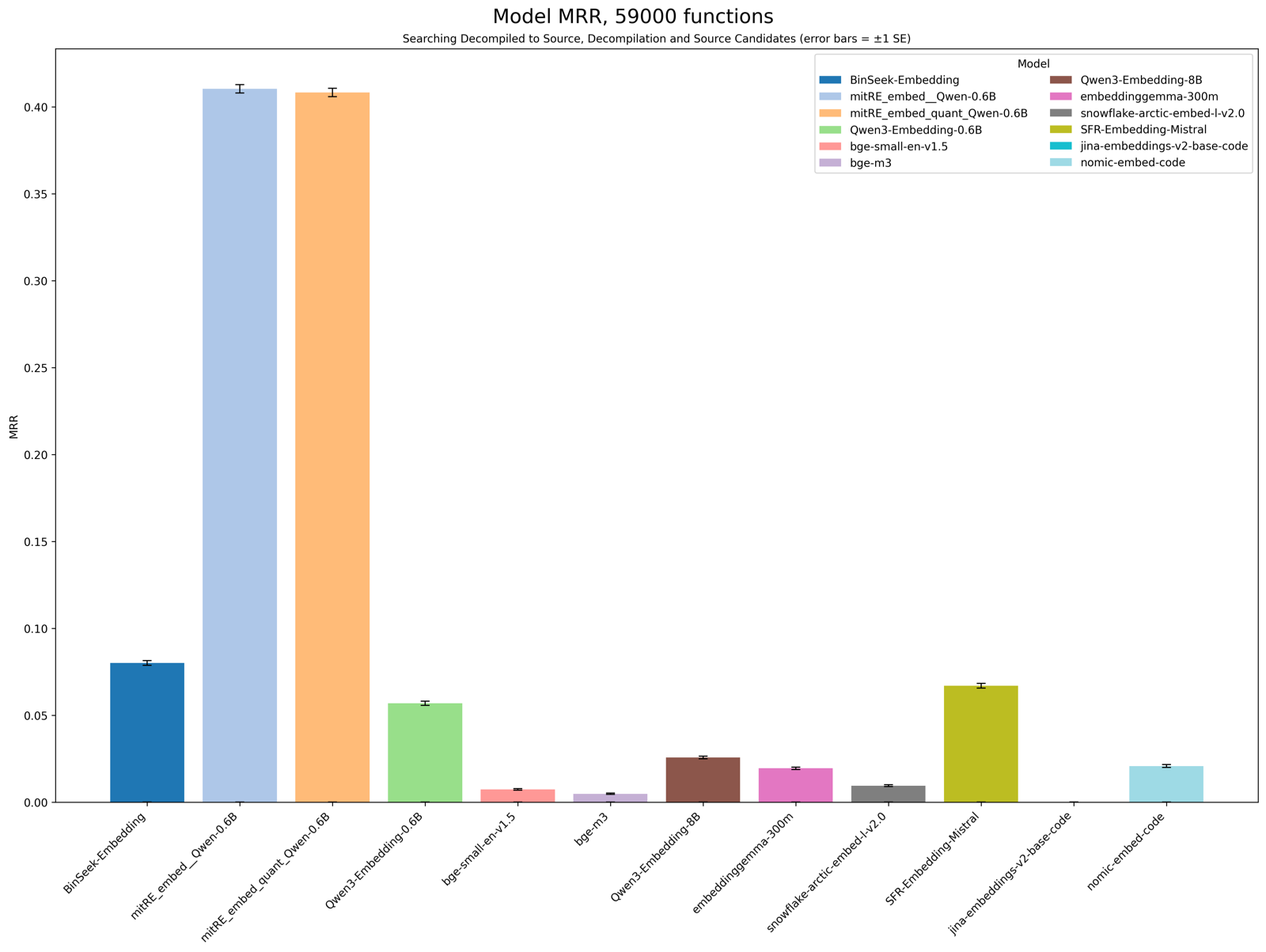}
    \caption{Assemblage MRR for \textbf{Decompiled \textrightarrow \space Source} with both representations in the search pool. Our model exhibits the best performance.}
    \label{fig:mrr_1}
\end{figure}

\begin{figure}[htbp]
    \centering
    \includegraphics[width=0.9\textwidth]{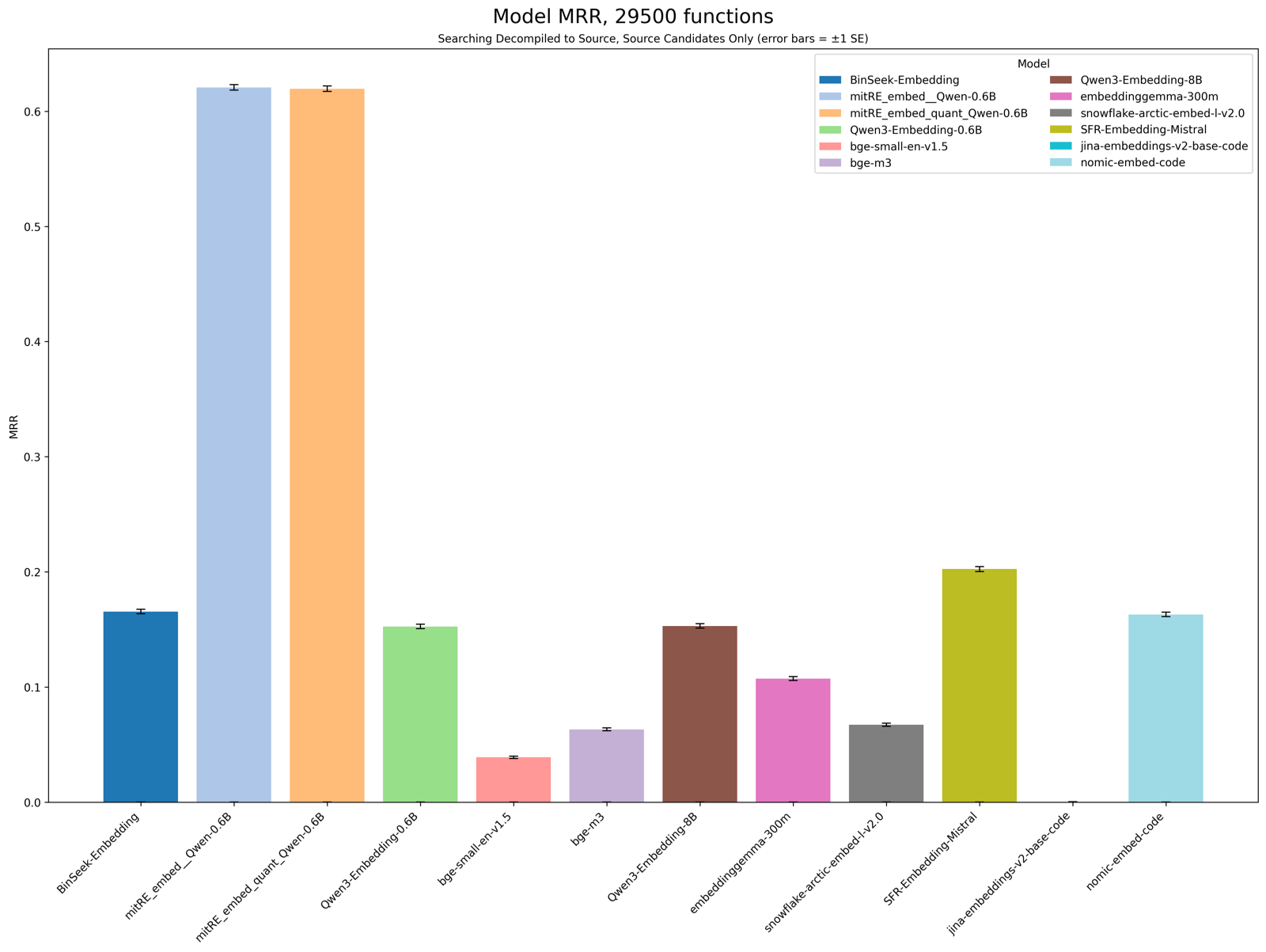}
    \caption{Assemblage MRR for \textbf{Decompiled \textrightarrow \space Source} with a filtered search pool. Our model exhibits the best performance.}
    \label{fig:mrr_2}
\end{figure}

\begin{figure}[htbp]
    \centering
    \includegraphics[width=0.9\textwidth]{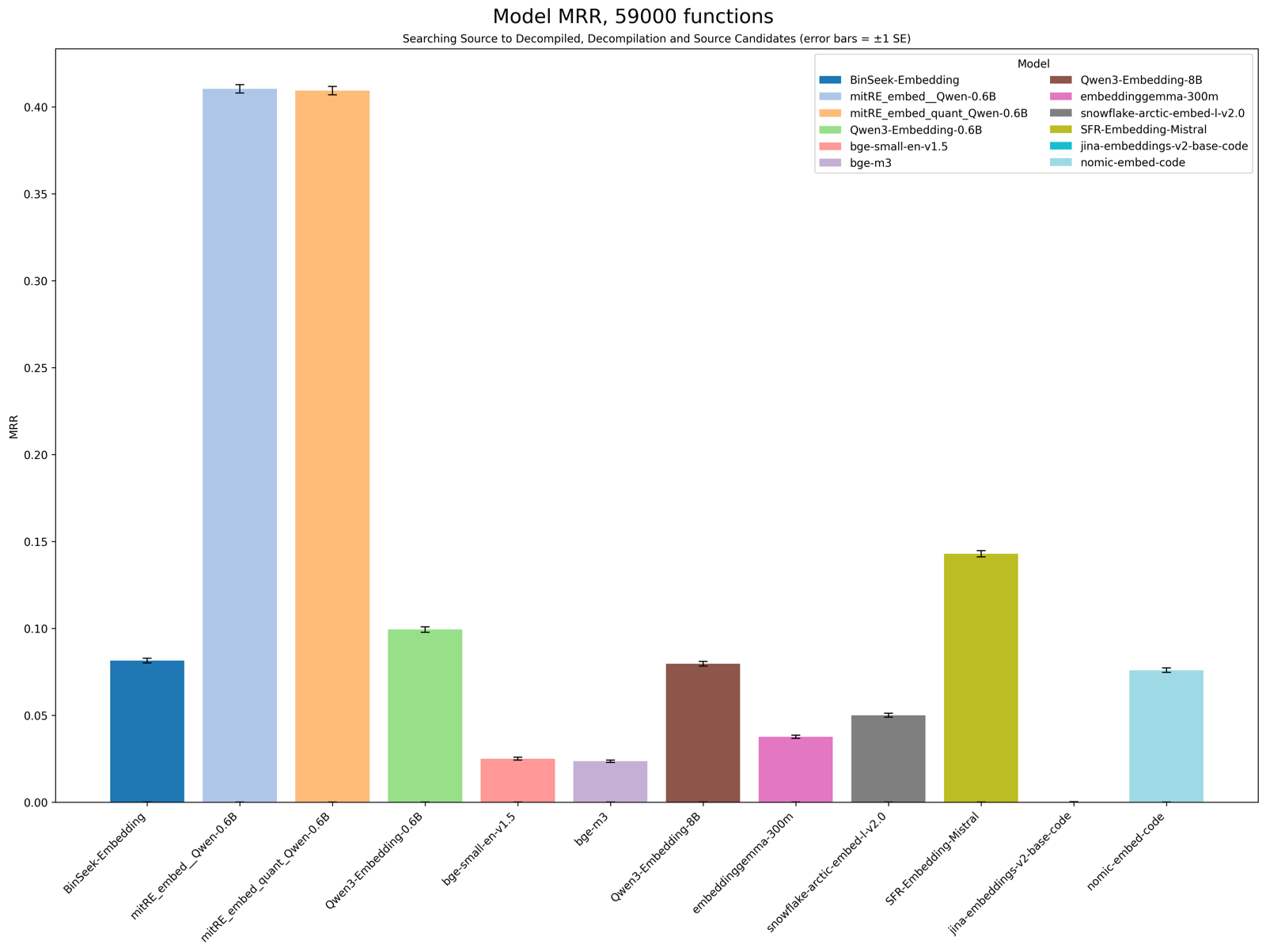}
    \caption{Assemblage MRR for \textbf{Source \textrightarrow \space Decompiled} with both representations in the search pool. Our model exhibits the best performance.}
    \label{fig:mrr_3}
\end{figure}

\begin{figure}[htbp]
    \centering
    \includegraphics[width=0.9\textwidth]{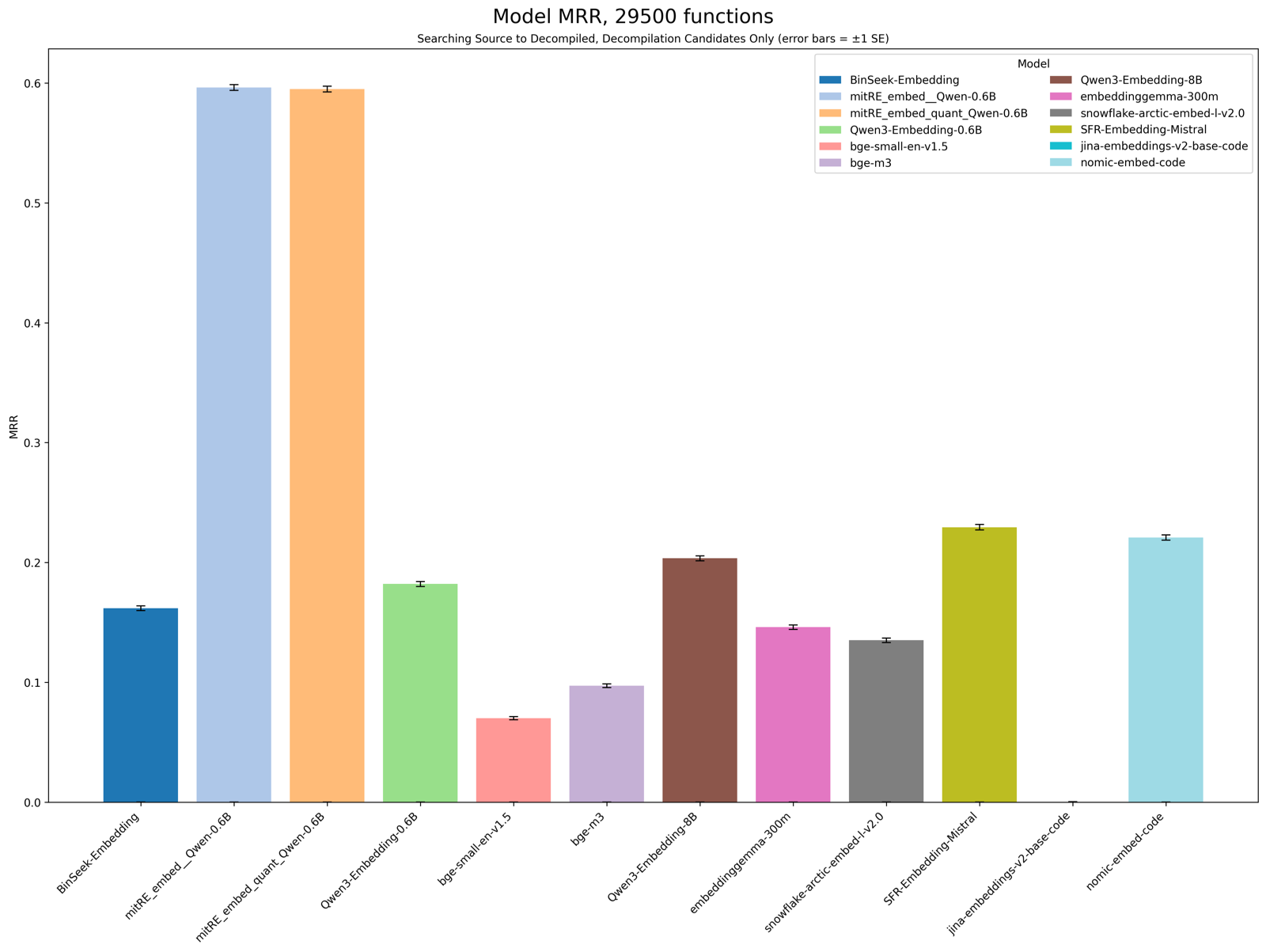}
    \caption{Assemblage MRR for \textbf{Source \textrightarrow \space Decompiled} with a filtered search pool. Our model exhibits the best performance.}
    \label{fig:mrr_4}
\end{figure}

\section{Signsrch Experiment Bar Graphs with Error}

We plot the mean AP and the mean MAP respectively for the Signsrch experiments with standard error bars in Figures 13 and 14. These standard error values are standard error of the mean and reflect variance across queries.

\begin{figure}[htbp]
    \centering
    \includegraphics[width=1\textwidth]{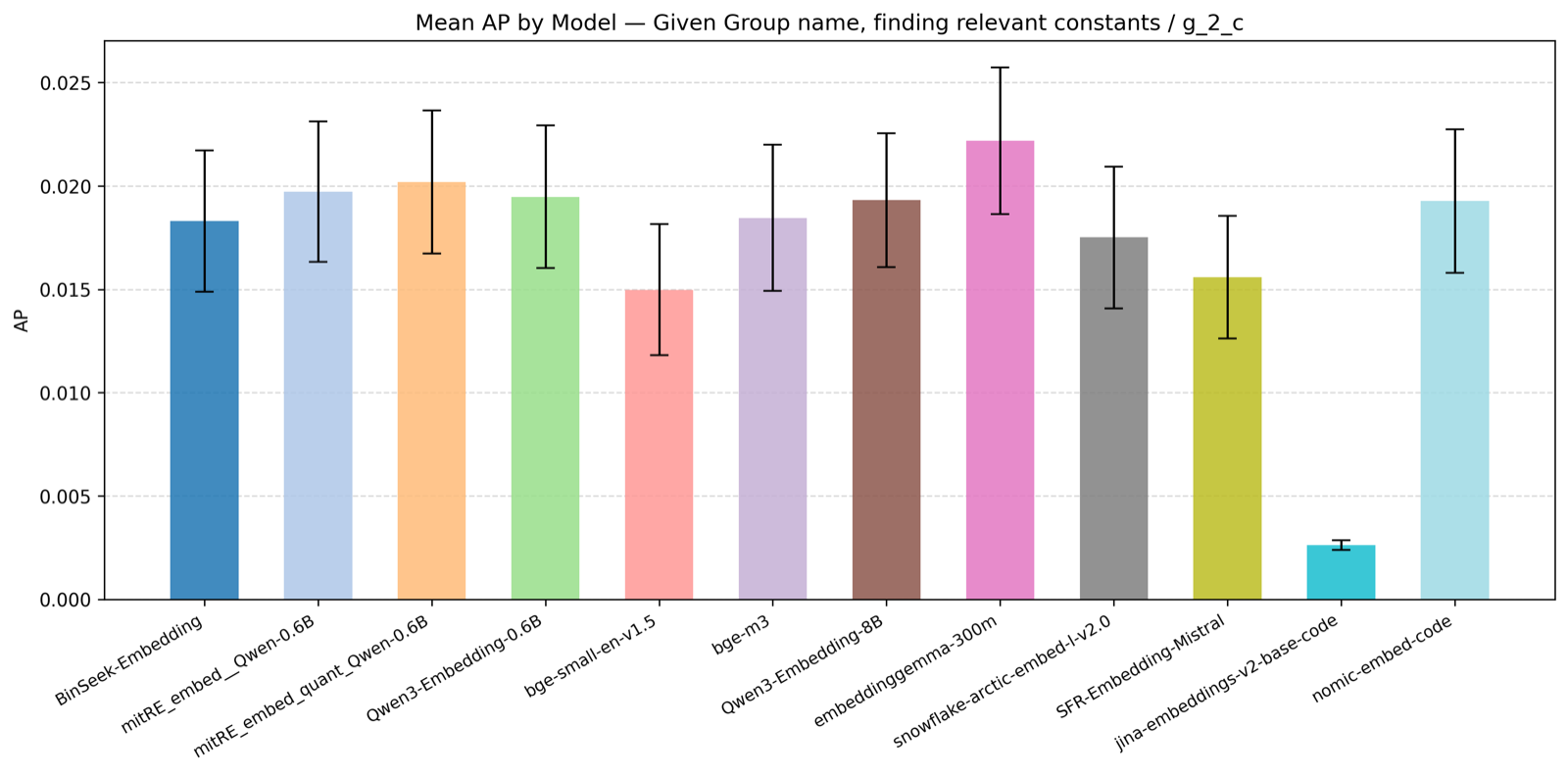}
    \caption{Signsrch results for \textbf{Algorithm Group Name \textrightarrow \space Constants}. Embeddinggemma-300m exhibits the best performance.}
    \label{fig:signsrch_bar_1}
\end{figure}

\begin{figure}[htbp]
    \centering
    \includegraphics[width=1\textwidth]{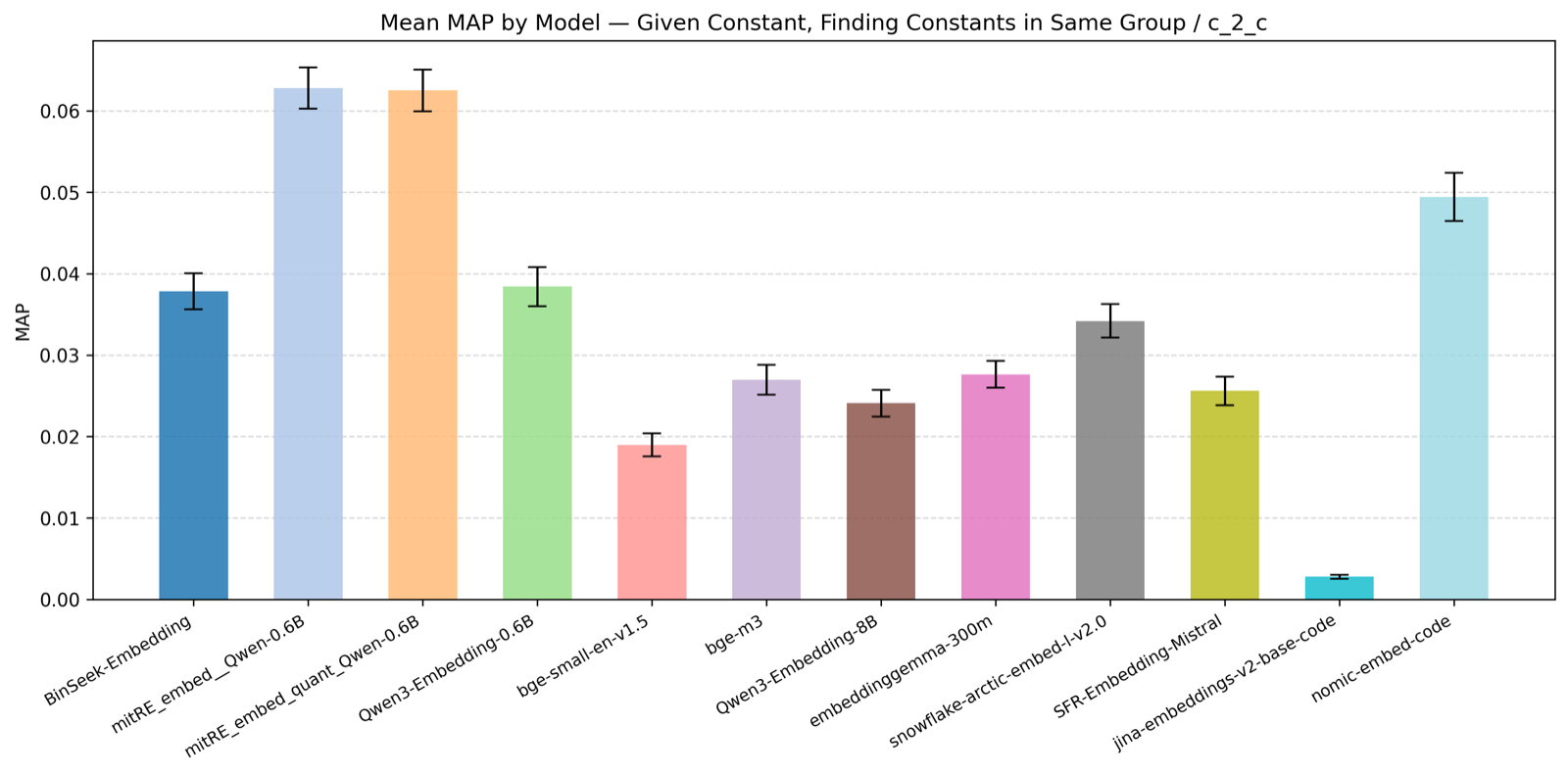}
    \caption{Signsrch results for \textbf{Constant \textrightarrow \space Other Group Constants}. Our model exhibits the best performance.}
    \label{fig:signsrch_bar_2}
\end{figure}


\end{document}